\documentclass{article}
\usepackage[twocolumn]{emulateapj}
\slugcomment{Accepted for publication in ApJ Letters}

\newcommand\etal{{ et al. }}
\def\lsim{\mathrel{\rlap{\lower 4pt \hbox{\hskip 1pt $\sim$}}\raise 1pt \hbox
        {$<$}}}
\def\gsim{\mathrel{\rlap{\lower 4pt \hbox{\hskip 1pt $\sim$}}\raise 1pt \hbox
        {$>$}}}

\begin{document}

\righthead{Impacts of the Detection of Cassiopeia A Point Source}
 
\title{Impacts of the Detection of Cassiopeia A Point Source}

\author{Hideyuki Umeda, Ken'ichi Nomoto}
\affil{Department of Astronomy and Research Center for the Early Universe, 
University of Tokyo, Bunkyo-ku, Tokyo 113-0033, Japan \\ e-mail: 
umeda@astron.s.u-tokyo.ac.jp, nomoto@astron.s.u-tokyo.ac.jp}

\author{Sachiko Tsuruta}
\affil{Department of
Physics, Montana State University, Bozeman, 
Montana 59717, USA 
\\ e-mail: 
uphst@gemini.oscs.montana.edu}

\author{Shin Mineshige}
\affil{Department of
Astronomy, Kyoto University, Sakyo-ku, Kyoto 606-8502, Japan
\\ e-mail: 
minesige@kusastro.kyoto-u.ac.jp}

\begin{abstract}

Very recently the {\it Chandra} First Light Observation discovered a
point-like source in Cassiopeia A (Cas A) supernova remnant.  This
detection was subsequently confirmed by the analyses of the archival
data from both ROSAT and {\it Einstein} observations.  Here we compare
the results from these observations with the scenarios involving both
black holes (BH) and neutron stars (NS).  If this point source is a BH
we offer as a promising model a disk-corona type model with low
accretion rate where a soft photon source at $\sim$ 0.1 keV is
Comptonized by higher energy electrons in the corona.  If it is a NS
the dominant radiation observed by {\it Chandra} most likely
originates from smaller, hotter regions of the stellar surface, but we
argue that it is still worthwhile to compare the cooler component from
the rest of the surface with cooling theories. We emphasize that the
detection of this point source itself should potentially provide {\it
enormous impacts} on the theories of supernova explosion, progenitor
scenario, compact remnant formation, accretion to compact objects, and
neutron star thermal evolution.

\end{abstract}


\section{Introduction}

Cassiopeia A (Cas A) is an interesting supernova (SN) remnant in
various aspects. The remnant is very young, about 320 years old.  This
ring-shaped (e.g. Holt \etal 1994) remnant is associated with jet like
structures (Fesen, Becker \& Blair 1987).  The observed abundances of
heavy elements are in good agreement with the yields of a massive star
(e.g., Hughes \etal 2000).  The overabundance of nitrogen found in
some knots (Fesen \etal 1987) implies that the progenitor was a
massive Wolf-Rayet star (WN type) which has lost most of its H-rich
envelope during the pre-SN evolution.  The SN was suggested to be
faint (Ashworth 1980), which implies that the progenitor was not a
red-supergiant possibly due to loss of its H-rich envelope.

Recently the ACIS on board the {\it Chandra} X-ray satellite observed
Cas A and found a point-like source (Tananbaum \etal 1999).
Subsequently, Aschenbach (1999) reported that the ROSAT/HRI image of
Cas A taken during 1995-1996 also shows the point-like source at the
similar location.  Very recently Pavlov \etal (2000) and Chakrabarty
\etal (2000) reported the results of their detailed analyses of the
Cas A point-source data from the {\it Chandra} observation.  These
authors convincingly argue that the observed point source should,
indeed, be a compact remnant of the SN explosion.  The single
power-law fit to the {\it Chandra} data by Pavlov \etal (2000) yields
the higher photon index $\Gamma$ and lower luminosity $L$ than those
observed from typical young pulsars.  The spectrum can be equally well
fit by thermal models.  The best fit for a one component blackbody
model yields the temperature $T^{\infty} = 6-8$ MK, the effective
radius $R_e = 0.20-0.45$ km, and the bolometric luminosity
$L^{\infty}=(1.4-1.9)\times 10^{33}$ erg s$^{-1}$.  (In this paper the
temperature $T^{\infty}$ and luminosity $L^{\infty}$ refer to the
values to be observed at infinity.)  Chakrabarty \etal (2000) obtained
similar results.  The size is too small for a 10 km radius neutron
star (NS), but it is consistent if the dominant emission comes from
localized hot spots.  Pavlov \etal (2000) find that the spectrum is
equally well fit by a two temperature thermal model with hydrogen
polar caps and the rest of the cooling NS surface composed of Fe.
These authors also analyzed the archival data from ROSAT and {\it
Einstein}, and report that the results are consistent with the {\it
Chandra} results within the 1 $\sigma$ level.  Their data analyses of
the point source showed no statistically significant variability (both
long and short time scale) over the {\it Einstein} - {\it Chandra}
period.  Chakrabarty \etal (2000) carried out detailed timing analysis
and report that the 3 $\sigma$ upper limit on the sinusoidal pulsed
fraction is $<$ 25\% for period $P >$ 100 ms, $<$ 35\% for $P>$ 5 ms,
and $<$ 50\% for $P >$ 1 ms.

We emphasize here that {\it the detection of the point source itself
is extremely important}, whether it turns out to be a neutron star
(NS) or a black hole (BH).  In this paper, therefore, we will consider
both cases.  Although the currently available data are not sufficient
to distinguish between these options, the most recently completed long
{\it Chandra} observation by S. Holt \etal (2000) and already planned
long XMM observations should be able to do so.  Therefore, we consider
that {\it it is extremely important and timely now, to discuss the
implications and offer some predictions for each case}.

\section{Accreting Black Hole}

 If the Cas A progenitor is more massive than $\sim 25M_\odot$ a BH
may be formed in the explosion (e.g., Ergma \& van den Heuvel 1998).
After formation the inner part of the ejected matter may fall back
onto the BH due to the presence of a deep gravitational potential well
or a reverse shock. The property of an accreting BH depends strongly
on whether or not an accretion disk is formed. Here we present plausible
BH scenarios based on the disk accretion model under the
following observational constraints (see, e.g., Pavlov et al. 2000):
(1) the single power-law X-ray luminosity of intermediate
    brightness, $L_{\rm x}$(0.1$-$5.0 keV) $= 
    (2-60) \times 10^{34}$ erg s$^{-1}$ 
    for distance $d$ = 3.4 kpc, which is much lower 
    than the Eddington luminosity, $L_{\rm EDD} \sim 7.5
    \times 10^{38} (M_{\rm BH}/3M_\odot)$ erg s$^{-1}$ for
    hydrogen-free matter,
(2) no significant variability being detected between the 
    {\it Einstein} and {\it Chandra} observations,
(3) large $F_{\rm x}/F_{\rm opt}$ ($\gsim 100$), and
(4) large power-law photon index, $\Gamma \sim 2.6 - 4.1$.

In our model, we assume that the fallback material has specific
angular momentum greater than $\sim (GM_{\rm BH}r_{\rm S})^{1/2}$
(where $r_{\rm S}$ is the Schwarzschild radius) and thus a fallback
disk is formed.  There is no efficient mechanism for angular-momentum
removal since the Cas A compact remnant is unlikely to have a binary
companion (\S4).  Then the disk evolution most likely obeys the
self-similar solution in which the total angular momentum within the
disk is kept constant (Pringle 1974; Mineshige, Nomoto, \& Shigeyama
1993).  This solution predicts that disk luminosity decays in a
power-law fashion after the disk is formed (Mineshige \etal 1997) as
$l\equiv L/L_{\rm EDD}
   \sim 10 ({M_{\rm fallback}/0.1M_\odot})
             (\alpha/0.1)^{-1.3}(t/320{\rm yr})^{-1.3}
             $ $({M_{\rm BH}/3M_\odot})^{-1.15}$,
where $M_{\rm fallback}$ is the amount of fallback material and
$\alpha$ is the viscosity parameter.  We should allow a factor of
$0.1-10$ changes depending on the distribution of matter and angular
momentum. In order for the black-hole accretion scenario to be
consistent with the observed $l \sim 10^{-4}$ at 320yr, the amount of
the fallback material should indeed be very small, $M_{\rm fallback}
\sim 10^{-6}M_\odot$.  Although the accretion models predict
luminosity decrease during the last 20 years from the {\it Einstein}
(in 1979) to the {\it Chandra} (in 1999) observations, it is small,
only about 10\% -- $(320/300)^{-1.3} \sim 0.90$.  Since the {\it
Einstein} observations include larger error bars, more than several
tens of $\%$, the luminosity drop of this level cannot be detected,
which is consistent with the lack of observed long range large scale
variability.

The luminosity of $\sim 10^{33}$ erg s$^{-1}$ is typical to Galactic
BH candidates (GBHC) during quiescence.  However, the constraint (3),
the large $F_{\rm x}/F_{\rm opt}$ ratio, rules out models that invoke
formation of a fallback disk whose properties are similar to those in
quiescent GBHC (Chakrabarty et al. 2000).  In the case of usual GBHC,
hydrogen-rich matter is continuously added to the disk from the binary
companion.  According to the disk-instability model for outbursts of
GBHC (Mineshige \& Wheeler 1989), a part of the transferred material
is accumulated in the outer parts of the disk, which inevitably
produces large optical flux in the quiescent GBHC.  Also the
constraint (4), a large photon index $\Gamma$, is in conflict with the
ADAF (advection-dominated) model for the quiescent GBHC (Narayan,
McClintock, \& Yi 1996).  For any ADAF models in which soft photons are
provided only by internal synchrotron emission and no external soft
photons are available, the power-law photon indices should be as small
as $\Gamma \sim 1.7$ (Tanaka \& Lewin 1995).  These are the reasons
why Chakrabarty et al. (2000) did not favor an accreting BH model for
the Cas A source.

 Here we propose a different promising BH model, a {\it disk-corona}
type model, for which the above analogy to the GBHC {\it is not
valid}.  First, we consider the constraint (3), the large $F_{\rm
x}/F_{\rm opt}$ ratio.  In our model for Cas A, there is no binary
companion which supplies mass at 320 years (\S4).  This means that the
outer disk boundary is not extended enough to emit significant optical
fluxes.  The disk is stable due to the smaller disk size and the
different composition of the disk material (mostly heavy elements with
possibly a little He but no hydrogen; \S4), i.e., the thermally
unstable outer zones are absent.  In the absence of an instability,
the mass-flow rate in the disk is close to be constant (Mineshige
\etal 1993).  Then according to the standard disk model, the effective
temperature is $T(r) \sim 4000 (M_{\rm BH}/3M_\odot)^{1/4}
(r/10^{10}{\rm cm})^{-3/4} ({l}/10^{-4})^{1/4}$ K.  For the disk size
as small as $r_{\rm d} \lsim 10^{10}$ cm and $l \sim 10^{-4}$, the
constraint $L_{\rm opt} < 10^{32}$ erg s$^{-1}$ is satisfied.

 Next consider the constraint (4), the large $\Gamma$.  In order to
reproduce large photon indices by Compton scattering, we require that
energy input rate into soft photons exceeds that into electrons.  It
is important to note that GBHC generally exhibit two states, soft and
hard, and a large $\Gamma$ is a characteristics of the soft-state
emission which exhibits soft blackbody spectra with $kT \sim 1$ keV.
The radiation from thermal photons with $\sim 1 $ keV times the area
of the emission region around a typical black hole of 3 -- 10
M$_\odot$ produces higher luminosity, $L_{\rm x} \sim 10^{36-38}$ erg
s$^{-1}$, than observed from Cas A.  However, we emphasize that unlike
GBHC no further mass input is available in our model.  Then the
accretion rate monotonically decreases, and so does the maximum
blackbody temperature, as $T_{\rm max} \sim 0.1
(M/3M_\odot)^{1/4}({l}/10^{-4})^{1/4}$ keV.  Therefore, we get $T_{\rm
max} \sim$ 0.1 keV for $l \sim 10^{-4}$, instead of $\sim$ 1 keV.  A
large $\Gamma$ is then naturally obtained in our model, since there is
copious supply of soft-photons at $\sim$ 0.1 keV into electron clouds
in the corona from the underlying cool disk (Mineshige, Kusunose,
Matsumoto 1995).  In other words, the important model parameter is
$\ell_{\rm soft}/\ell_{\rm hard}$ (ratio of compactness parameter of
soft photons to that of hard electrons), where the compactness
parameter is proportional to the energy output rate divided by the
size of the region.  For $\ell_{\rm soft} > \ell_{\rm hard} $, we have
a large spectral index ($\Gamma >$ 2) because of efficient Compton cooling of
hard electrons as shown in Mineshige et al. (1995).  The spectral
slope is rather insensitive to $\dot M$ and $M$.  The conclusion is
that with the low accretion rate and lower soft photon temperature our
Compton model with a disk-corona configuration naturally yields large
$\Gamma$ with the observed luminosity.

\section{Cooling Neutron Star}

Here let us assume that the observed Cas A point source is a NS.
Pavlov \etal (2000) and Chakrabarty \etal (2000) convincingly argue
that the dominant radiation observed by {\it Chandra} is most likely
coming from polar hot spots or equatorial ring if it is a NS.  Our
main purpose in this section is to argue that it is still worthwhile
to compare with theoretical models the observed upper limit to the
cooling NS component (i.e., the radiation from the whole stellar
surface excluding the hotter, localized areas).

Pavlov \etal (2000) offered, as a possible model, a two-component
thermal model where the temperature and radius of the polar caps with
hydrogen are 2.8 MK and $\sim$ 1 km, respectively, while the rest of
the surface of the 10 km NS consisting of Fe is at 1.7 MK.  In this
model, the hotter polar caps are the result of higher conductivity of
hydrogen as compared with Fe, the temperature difference between the
polar caps and the rest of the surface should be small, less than a
factor of 2, and hence non-standard cooling should be excluded.

Here we offer a promising alternative NS model.  SN remnants are
usually classified into two categories: shell-type and filled-center
(plerions). Cas A is considered to be a prototype of the former, where
radio pulsars are normally not found.  Recently Pacini (2000)
emphasized the evidence for the presence of an active NS in at least
some of the shell-type SNRs although radio pulsars were not found.
Also, there is some evidence for significant magnetospheric activities
(which can be responsible for polar cap heating) in some NS where no
radio pulsar has been found.  An example is Geminga (e.g., see Tsuruta
1998).  Therefore, the apparent absence of a radio pulsar and/or a
plerion should not be used as evidence against polar cap heating.
Chakrabarty \etal (2000) offers accretion as a possible cause for
polar cap heating when the field strength is significant.  If it is
weak, their accreting NS model offers the hotter component as
originating from the equatorial hot ring.  In either case, with an
additional heat source for the hotter component, larger temperature
difference between the hotter and cooler components is expected, and
hence there is no conflict with the possibility of faster non-standard
cooling.

We adopt the conservative upper limit to the cooler component given by
{\it Chandara} (Pavlov \etal 2000), $L^{\infty} < 3 \times 10^{34}$
erg s$^{-1}$.  The neutron star thermal evolution is calculated with a
general relativistic evolutionary code without making the isothermal
approximation (Nomoto \& Tsuruta 1987, Umeda \etal 1994,
hereafter U94; Umeda, Tsuruta \& Nomoto 1994, hereafter UTN94).  Our
results are summarized in Figure 1.

%\placefigure{FIG1}
 
The observed upper limit for Cas A is consistent with the 'standard'
cooling.  However, it is still only an upper limit, and if
the actual luminosity of the cooler component turns out to be $\sim
10^{33}$ erg s$^{-1}$ or less, the result will be extremely
interesting.  This is because then the observed value will be
certainly below the standard cooling curve, and hence that will be
considered the evidence for non-standard cooling scenarios such as
those involving pion and/or kaon condensates, or the direct URCA
process (e.g., U94; UTN94). 
When the particles in the stellar core are in the
superfluid state with substantial superfluid energy gaps, neutrino
emissivity $l_\nu$ is significantly suppressed (e.g., see Tsuruta
1998).
In order to examine this effect of superfluidity, we calculated
pion cooling for a representative superfluid model with an
intermediate degree of suppression, called the E1 -- 0.6 model (see
U94). The result is shown as the thin solid curve in Figure 1.

\section{Constraints from Progenitor Scenarios}

Here we discuss whether the formation of a NS or BH is consistent with
the current models of stellar evolution and supernovae and whether the
evolutionary scenarios constrain the radiation processes from the
compact source.  The overabundance of nitrogen in Cas A implies that
the progenitor was a massive WN star which lost most of its hydrogen
envelope before the SN explosion. Here we describe two possible
evolutionary paths to form such a pre-SN WN star.

 One path is the mass loss of a very massive {\it single} star.  A
star with the zero-age main-sequence mass $M_{\rm MS}$ larger than
$\sim$ 40 M$_\odot$ can lose its hydrogen-rich envelope via mass loss
due to strong winds and become a Wolf-Rayet star (e.g., Schaller \etal
1992).  Recent theoretical models and population synthesis studies
suggest that stars with $M_{\rm MS} \gsim $ 25 M$_\odot$ are more
likely to form BHs than NSs (e.g., Ergma \& van den Heuvel 1998).
This implies that the WN star progenitor is massive enough to form a
BH.  The explosion can be energetic enough to prevent too much matter
fall back to be consistent with the small fall back mass inferred in
\S2.

 The other evolutionary path to form a pre-SN WN star is mass loss due
to binary interaction.  If the progenitor is in a close binary system
with a less massive companion star, the star loses most of its H-rich
envelope through Roche lobe overflow.  In this case, the WN progenitor
can form from a star of $M_{\rm MS} \lsim 40$ M$_\odot$.  Its SN
explosion of type Ib/c would leave either a BH (if $M_{\rm MS} \sim
25-40$ M$_\odot$) or a NS (if $M_{\rm MS} \lsim 25$ M$_\odot$).  If
the compact remnant in Cas A turns out to be a NS, therefore, the
progenitor must have been in a close binary system.

In the binary scenario, the companion to the Cas A progenitor cannot
be more massive than a red dwarf, as constrained from the R \& I band
magnitude limit (van den Bergh \& Pritchet 1986). When the companion
star is such a small mass star, i.e., the mass ratio between the stars
is large, the mass transfer is inevitably non-conservative (e.g.,
Nomoto, Iwamoto, \& Suzuki 1995), and the companion star will
spiral-in into the envelope of the Cas A progenitor.  In order for
most of the H-rich envelope to be removed, the envelope should have
been a red-giant size so that the orbital energy released during the
spiral-in exceeds the binding energy of the envelope.  After losing
its envelope due to frictional heating, the star became a WN star.

If we take the model of $M_{\rm MS} = 25 M_\odot$, as an example, the
star at the WN stage has 8 $M_\odot$.  Since the explosion ejects Si
and Fe from the deep layers (Hughes et al. 2000), the mass of the
compact remnants could not exceed $2 - 3 M_\odot$.  Then the binary
system is very likely to be disrupted at the explosion.  Then the
compact star in the Cas A remnant does not have a companion star, and
so no mass transfer can be postulated.  The implication is also that
the accretion onto the compact remnant can occur only as a result of
fallback of the ejected matter, and so the composition of the fall
back matter is mostly heavy elements with possibly a small fraction of
helium but no hydrogen.

In either the single or binary scenario, the WN star blows a fast wind
which collides with the red-giant wind material to form a dense shell
(Chevalier \& Liang 1989).  If the red-giant wind formed a ring-like
shell (due possibly to the spiral-in of the companion), the collision
between the supernova ejecta and the shell could explain the observed
ring-like structure of Cas A.

\section{Discussion and Conclusion}

We agree with Chakrabarty et al. (2000) that for Cas A point source
the usual ADAF model for a quiescent GBHC hardly reconcile with
observation. However, we emphasized in \S2 that there does exist a
very promising BH disk accretion model.  In this model, the fallback
material is like the soft state of a GBHC with a disk-corona
configuration, not like a quiescent GBHC with ADAF.  With the low
accretion rate and Comptonization of cooler soft photons ($\sim$ 0.1
keV or less), we naturally obtain large photon index of $\Gamma \sim
2.6-4.1$ and lower luminosity of $L \sim 10^{34}-10^{35}$ erg
s$^{-1}$, as observed from the Cas A point source.

Accreting NS models are also possible (see Chakrabarty \etal 2000).
However, we can still, without difficulty, distinguish between the BH
and NS accretion models because the characteristic properties of the
observed X-ray spectra in these two cases are quite different (e.g.,
see Tanaka 2000).  For instance, the radiation from an accreting NS is
dominated by thermal emission from the stellar surface (Rutledge \etal
2000), which is absent if a BH is involved.

If the point source is a NS, the dominant radiation observed by {\it
Chandra} most likely corresponds to the radiation from a localized
small area.  The detailed studies of theoretical light curves expected
 from anisotropic cooling of a NS have been carried out by, e.g.,
Shibanov \etal (1995) and Tsuruta (1998), with the latter including
hot spots.  The results show that pulsation depends on the relative
angles between the rotation axis, magnetic axis and the line of sight.
Depending on the combinations of these angles, pulsations from zero to
up to about 30\% are predicted, and so the observed constraints on the
pulsed fraction are still consistent with a NS model.

Although the current data of Cas A point-source can be consistent with
both BH and NS scenarios, future observations by the {\it Chandra},
XMM, and other satellite missions should be able to distinguish
between these cases. If distinct periodicity is found the point source
definitely should be a NS.  The existence of the NS itself will
significantly constrain the progenitor scenario for Cas A.  Better
spectral information should be able to distinguish between the BH and
NS as the compact remnant. If the source is found to be a BH, the
implication is significant in the sense that this will offer the first
observational evidence for BH formation through a SN explosion and
greatly constrain the BH progenitor mass by combining with the
abundance analysis of Cas A (Hughes et al. 2000).

In conclusion we emphasize that the Cas A point source can potentially
provide great impacts on the theories of supernova explosion,
progenitor scenario, compact remnant formation, accretion to compact
objects, and NS thermal evolution.
 
We thank Drs. G. Pavlov, M. Rees, H. Tananbaum, B. Aschenbach, and
J. Tr\"umper for valuable discussions.  This work has been supported
in part by the grant-in-Aid for Scientific Research (0980203,
09640325), COE research (07CE2002) of the Ministry of Education,
Science, Culture and Sports in Japan, and a NASA grant NAG5-3159.

\clearpage

\begin{figure}
\epsscale{.5}
\plotone{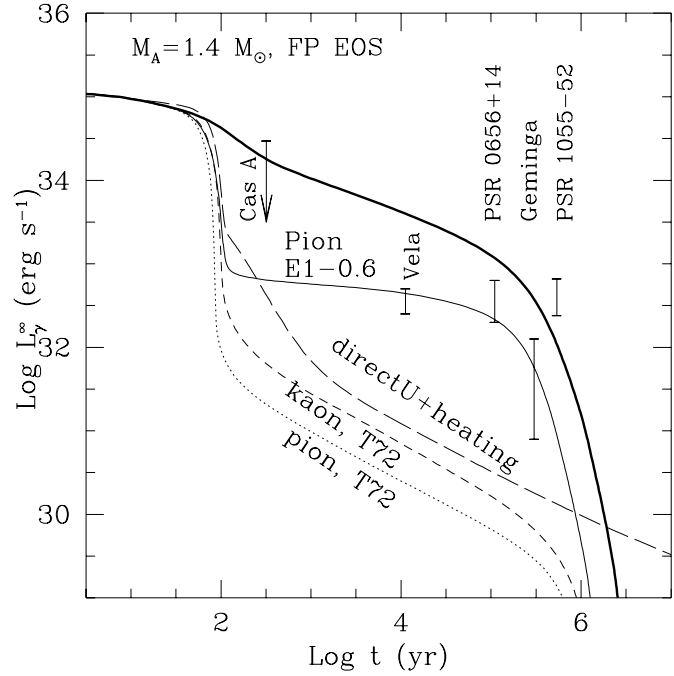}
\caption{
 Various neutron star thermal evolution curves are compared with the
observational data of the Cas A point source and several cooling NS
candidates.  The heavy solid curve refers to `standard-cooling', the
dashed and dotted curves show the `non-standard' kaon and pion cooling
scenarios when the superfluid effect is negligible, while the thin
sold curve refers to the pion cooling with significant superfluid
effect.  The long dashed curve shows the effect of strong heating on
the direct URCA cooling, the fastest `non-standard' scenario.  These
curves are obtained for 1.4 $M_\odot$ neutron stars with the
intermediate FP equation of state (Friedman \& Pandharipande
1981). Cooling due to various more straightforward neutrino mechanisms
such as the modified URCA is called `standard cooling', while
extremely fast cooling caused by some other more unconventional
mechanisms is called `non-standard cooling'.  The Cas A data should be
considered as the upper limit to the radiation from the whole stellar
surface.
\label{FIG1}}
\end{figure}


\begin{thebibliography}{}

\bibitem[]{}

\bibitem[]{}
Aschenbach, B. 1999, IAUC No. 7249

\bibitem[]{}
Ashworth, W.B. 1980, J. Hist. Astron., 11, 1

\bibitem[]{}
Chakrabarty, D., Pivovaroff, M.J., Hernquist, L.E., Heyl, J.S., \&
Narayan, R. 2000, ApJ, submitted (astro-ph/0001026)

\bibitem[]{}
Chevalier, R.A., \& Liang, E. 1989, ApJ, 346, 847

\bibitem[]{}
Ergma, E., \&  van den Heuvel, E.P.J. 1998, A\&A, 331, L29 

\bibitem[]{}
Fesen, R.A., Becker, R.H., \& Blair, R.H. 1987, ApJ, 313, 378

\bibitem[]{}
Friedman, B., \& Pandharipande, V.R. 1981, Nucl. Phys. A, 361, 502

\bibitem[]{}
Holt, S., Gotthelf, E.V., Tsunemi, H., \& Negoro, H. 1994,
PASJ, 46, L151

\bibitem[]{}
Holt, S. et al. 2000,
 
http://asc.harvard.edu/targets/summary\_observed\_daily.html

\bibitem[]{}
Hughes, J.P., Rakowski, C.E., Burrows, D.N., \& Slane, P.O. 2000, ApJ,
528, L109

\bibitem[]{}
Mineshige, S., Kusunose, M., \& Matsumoto, R.. 1995, ApJ, 445, L43

\bibitem[]{}
Mineshige, S., Nomoto, K., \& Shigeyama, T. 1993, A\&A, 267, 95

\bibitem[]{}
Mineshige, S., Nomura, H., Hirose, M., Nomoto, K.,
\& Suzuki, T. 1997, ApJ, 489, 22

\bibitem[]{}
Mineshige, S., \& Wheeler, J.C. 1989, ApJ, 343, 241

\bibitem[]{}
Narayan, R., McClintock, J.E., \& Yi, I. 1996, ApJ, 457, 821

\bibitem[]{}
Nomoto, K., Iwamoto, K., \& Suzuki, T. 1995, Phys. Rep., 256, 173

\bibitem[]{}
Nomoto, K., \& Tsuruta, S. 1987, ApJ, 312, 711 

\bibitem[]{}
Pacini, F. 2000, in IAU Symp. 195, Highly Energetic Physical Processes
and Mechanisms for Emission from Astrophysical Plasmas, eds. P. Martens,
S. Tsuruta, \& M. Weber, PASP, in press

\bibitem[]{}
Pavlov, G.G., Zavlin, V.E., Aschenbach, B., \& Tr\"umper, J..E. 2000,
ApJ, 531, L53

\bibitem[]{}
Pringle, J.E. 1974, Ph.D. Thesis, University of Cambridge

\bibitem[]{}
Rutledge, R.E., Bildsten, L., Brown, E.F., Pavlov, G.G., \& Zavlin,
 V.E. 2000, ApJ, 529, 985

\bibitem[]{}
Schaller, G., Schaerer, D., Meynet, G., \& Maeder, A. 1992,
A\&AS, 96, 269

\bibitem[]{}
Shibanov, Yu.A., Zavlin, V.E., Pavlov, G., Qin, L., \&
Tsuruta, S. 1995, Proc. 17th Texas Symp., eds. H. B\"ohringer, et al., 
N.Y. Acad. Sci., 291.

\bibitem[]{}
Tanaka, Y. 2000, in IAU Symp. 195, Highly Energetic Physical Processes
and Mechanisms for Emission from Astrophysical Plasmas, eds. P. Martens, 
S. Tsuruta, \& M. Weber, PASP, in press

\bibitem[]{}
Tanaka, Y., \& Lewin, W. H. G., 1995, in X-ray binaries,
ed. W.H.G. Lewin, J. van Paradijs, E.P.J. van den Heuvel (Cambridge:
U.P., Cambridge), 126

\bibitem[]{}
Tananbaum, H. \etal 1999, IAUC No. 7246

\bibitem[]{}
Tsuruta, S. 1998, Phys.. Rep. 292, 1

\bibitem[]{}
Umeda, H., Nomoto, K., Tsuruta, S., Muto, T., \&
Tatsumi, T. 1994, ApJ, 431, 309 (U94)

\bibitem[]{}
Umeda, H., Tsuruta, S., \& Nomoto, K. 1994, ApJ, 433, 256 (UTN94)

\bibitem[]{}
van den Bergh, S., \& Pritchet, C.J. 1986, ApJ, 307, 723

\end{thebibliography}
\end{document}